\documentclass[a4paper,11pt]{article}
\usepackage{pos}
\title{Precision bottomonium properties and b quark mass from lattice QCD}

\manuallySeparateAuthors
\author*[a]{C. T. H. Davies}
\author[a]{, D. Hatton}
\author[b]{, J. Koponen }
\author[c]{, G. P. Lepage}
\author[d]{ and A. T. Lytle }
\author{ for the HPQCD collaboration}

\affiliation[a]{SUPA, School of Physics and Astronomy, University of Glasgow, \\Glasgow, G12 8QQ, UK}

\affiliation[b]{PRISMA+ Cluster of Excellence and Institut f\"{u}r Kernphysik, Johannes Gutenberg-Universit\"{a}t Mainz, \\D-55128 Mainz, Germany} 

\affiliation[c]{Laboratory for Elementary-Particle Physics, Cornell University, \\Ithaca, NY 14853, USA} 

\affiliation[d]{Department of Physics, University of Illinois, \\Urbana, IL 61801, USA}

\emailAdd{christine.davies@glasgow.ac.uk}
\emailAdd{daniel.hatton@glasgow.ac.uk}
\emailAdd{jkoponen@uni-mainz.de}
\emailAdd{atlytle@illinois.edu}
\emailAdd{gpl3@cornell.edu}

\abstract{As tests of QCD in the bottomonium system, we give the most accurate results to date for the ground-state hyperfine splitting and the $\Upsilon$ leptonic width from full lattice QCD. These quantities are both accurately known from experiment, so can provide a good test of $b$ physics, but previous lattice results have been rather imprecise. We also test the impact on these quantities of the $b$ quark’s electric charge. Our results are: $M_{\Upsilon} -M_{\eta_b} = 57.5(2.3)(1.0) \,\mathrm{MeV}$ (where 
the second uncertainty comes from neglect of quark-line disconnected correlation 
functions) and decay constants, $f_{\eta_b}=724(12)$ MeV and 
$f_{\Upsilon} =677.2(9.7)$ MeV, giving 
$\Gamma(\Upsilon \rightarrow e^+e^-) = 1.292(37)(3) \,\mathrm{keV}$. 
	We also give a new determination of the ratio of the masses for $b$ and $c$ quarks that is completely nonperturbative in lattice QCD and includes the calculation of QED effects for the first time. This gives a result for the $b$ quark mass of $\overline{m}_b(\overline{m}_b,n_f=5) = 4.202(21) $ GeV. }

\FullConference{%
 The 38th International Symposium on Lattice Field Theory, LATTICE2021
  26th-30th July, 2021
  Zoom/Gather@Massachusetts Institute of Technology
}

\begin{document}
\maketitle

\section{Introduction}
Lattice QCD calculations of weak decay matrix elements for hadrons containing $b$ quarks are critical to the flavour physics programme. It is therefore important to have stringent tests of lattice QCD results for $b$ physics in other settings to make sure systematic errors are under control. Here we provide such tests in the bottomonium system with the ground-state hyperfine splitting and the $\Upsilon$ leptonic width, both accurately known from experiment. We give the most accurate lattice QCD results to date for these quantities and also test the impact on them of the $b$ quark’s electric charge~\cite{Hatton:2021dvg}.
	Accurate masses for heavy quarks are important for high-precision searches for new physics in Higgs decay~\cite{Lepage:2014fla}. We give here a new determination of the ratio of the masses for $b$ and $c$ quarks that is completely nonperturbative in lattice QCD and includes the calculation of QED effects for the first time~\cite{Hatton:2021syc}. This allows an accurate determination of the $b$ quark mass using our earlier accurate results for $m_c$ in QCD+QED~\cite{Hatton:2020qhk}. 

\section{Lattice QCD calculation}
We use the Highly Improved Staggered Quark (HISQ) action~\cite{hisqdef} on ensembles of gluon field configurations that include 2+1+1 flavours of HISQ quarks in the sea, generated by the MILC collaboration~\cite{Bazavov:2012xda}. The ensembles have a range of lattice spacing values from 0.09 fm down to 0.03 fm and with $u/d$ sea masses varying from 1/5 that of the $s$ quark ($M_{\pi} \approx$ 300 MeV) down to their physical value. On each ensemble we calculate pseudoscalar and vector 2-point correlators for HISQ valence quarks with masses $m_h$ from that of the $c$ quark up to $am_h=$ 0.8 or 0.9. On the finest lattices we can reach the $b$ quark mass with $am_h=0.65$. We calculate correlators including quenched QED on a subset of ensembles. Note that we do not include quark-line disconnected correlators. More details are given in~\cite{Hatton:2021dvg}. 

We determine the masses and amplitudes of the ground state mesons (denoted $\eta_h$ and $\phi_h$) on each ensemble using standard correlator fits. The hyperfine splitting is the difference of masses, $M_{\phi}-M_{\eta}$. The vector and pseudoscalar decay constants are determined from the amplitudes in the standard way; both are normalised accurately using lattice Ward identities~\cite{Hatton:2019gha}. We then perform a model-independent fit to these results as a function of lattice spacing and $\phi_h$ mass, using cubic splines.  This allows us to determine results in the continuum limit with physical sea quark masses for the case where the heavy quark is the $b$ quark, i.e. where the $\phi_h$ has the experimental mass of the $\Upsilon$. We use this criterion for tuning the $b$ quark mass both in pure QCD and in QCD+QED. 

The ratio of quark masses is scheme and scale independent in pure QCD, but not in QCD+QED if the quarks have different electric charges. We calculate the ratio of $b$ and $c$ masses at 3 GeV in the $\overline{\text{MS}}$ scheme using a 3 step procedure (for full details see~\cite{Hatton:2021syc}). The first step is to fit results for $m_h/m_c$ as a function of the $\phi_h$ mass and lattice spacing in pure QCD. $m_c$ here is the tuned $c$ quark mass, obtained on each ensemble using the experimental $J/\psi$ meson mass~\cite{Hatton:2020qhk}. Evaluating the fit function in the physical-continuum limit at the $\Upsilon$ mass then gives the pure QCD $m_b/m_c$ value. The second step is to calculate how much this ratio changes if BOTH the $b$ and $c$ quarks have charge $Q=e/3$ so that the ratio is still scale-invariant. The third step corrects the final ratio in the $\overline{\text{MS}}$ scheme at 3 GeV to take the $c$ quark charge to 2e/3. This is the largest QED effect and is calculated from the results of~\cite{Hatton:2020qhk}. 

\section{Results - hyperfine splitting}

\begin{figure}[thb]
  \centering
  \includegraphics[width=0.45\hsize]{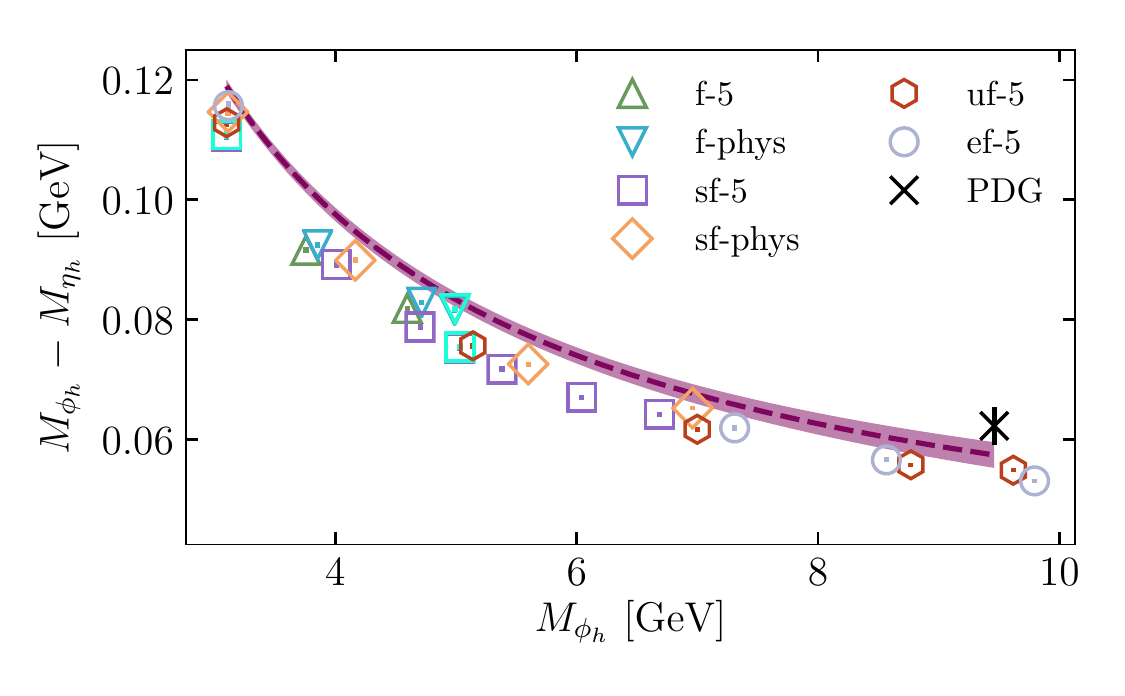}
    \includegraphics[width=0.45\hsize]{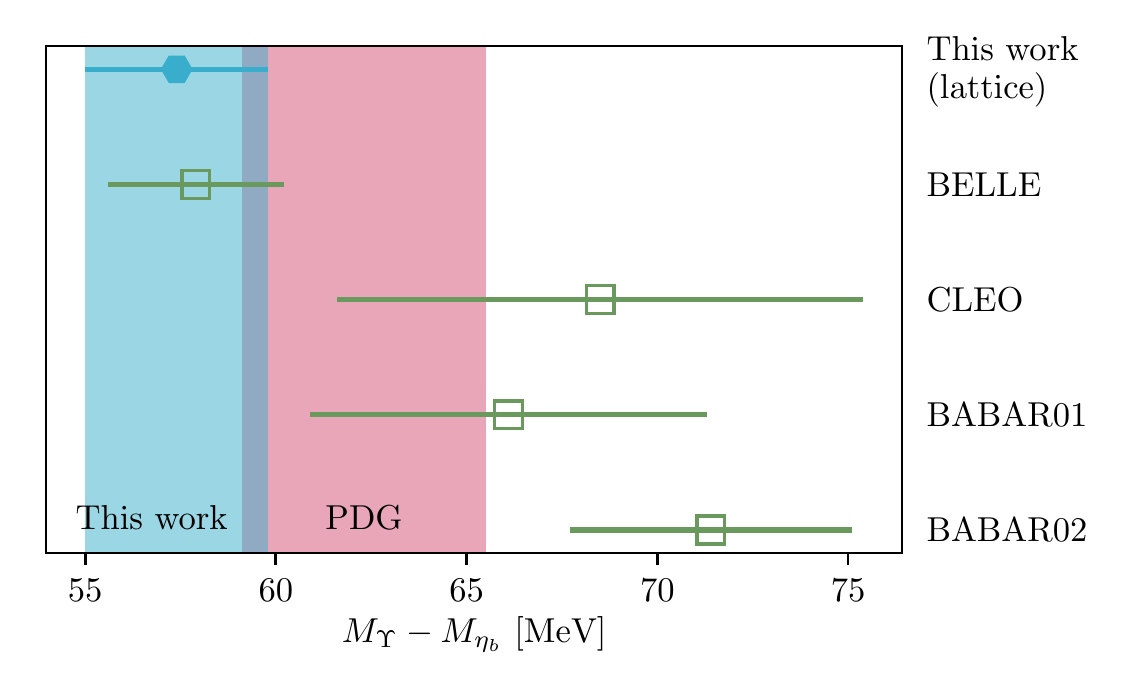}
  \caption{The left-hand plot shows the heavyonium hyperfine splitting as a function of the vector heavyonium mass from $n_f=2+1+1$ lattice QCD~\cite{Hatton:2021dvg}. The range covered is from the $c$ quark mass to the $b$ quark mass. Points show lattice QCD results at a range of lattice spacings from 0.09 fm (denoted $\text{f}$), through 0.06 fm (denoted $\text{sf}$) and 0.045 fm (denoted $\text{uf}$) to 0.03 fm (denoted $\text{ef}$). Sea $u/d$ quark masses are either $1/5$ of the $s$ quark mass (for points denoted $-5$) or physical. The purple band shows a model-independent fit to these results using cubic splines. The black cross marks the experimental average value~\cite{Zyla:2020zbs} for the ground-state bottomonium hyperfine splitting. The right-hand plot compares our result (denoted `This work') to recent experimental values. The pink band shows the experimental average~\cite{Zyla:2020zbs} and the blue band carries our result down the plot. Note the particularly good agreement between our value and the Belle result~\cite{Belle:2012fkf}, although we also agree well with the experimental average that includes in addition results from CLEO~\cite{CLEO:2009nxu} and BaBar~\cite{BaBar:2009xir, BaBar:2008dae}. }
  \label{fig:hyp}
\end{figure}

Figure~\ref{fig:hyp} shows our results~\cite{Hatton:2021dvg} for the hyperfine splitting as a function of $\phi_h$  mass along with our fit curve.  Our result at the $b$, i.e.\ the mass difference between $\Upsilon$ and $\eta_b$, is 57.5(2.3)(1.0) MeV, where the second uncertainty comes from the neglect of quark-line disconnected diagrams. Our result is compared to experimental results in the right-hand plot in the Figure. We see good agreement, particularly with the most recent result from Belle.  QED effects here (and in the decay constant below) are tiny.  

\section{Results - decay constants}

\begin{figure}[thb]
  \centering
  \includegraphics[width=0.45\hsize]{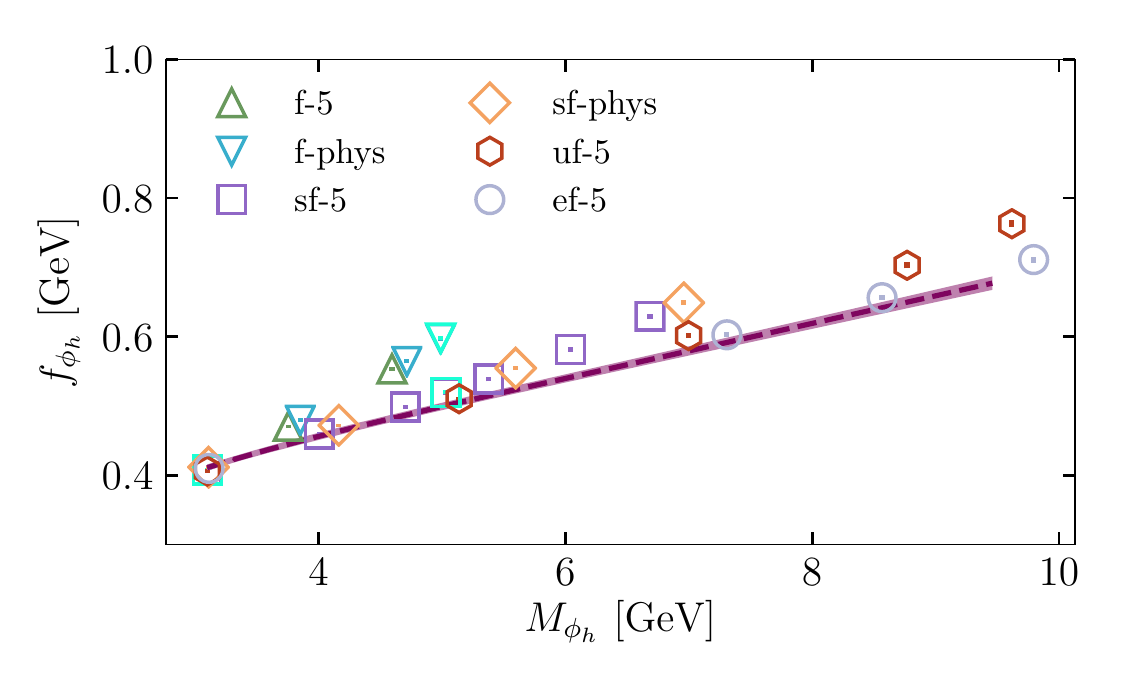}
    \includegraphics[width=0.45\hsize]{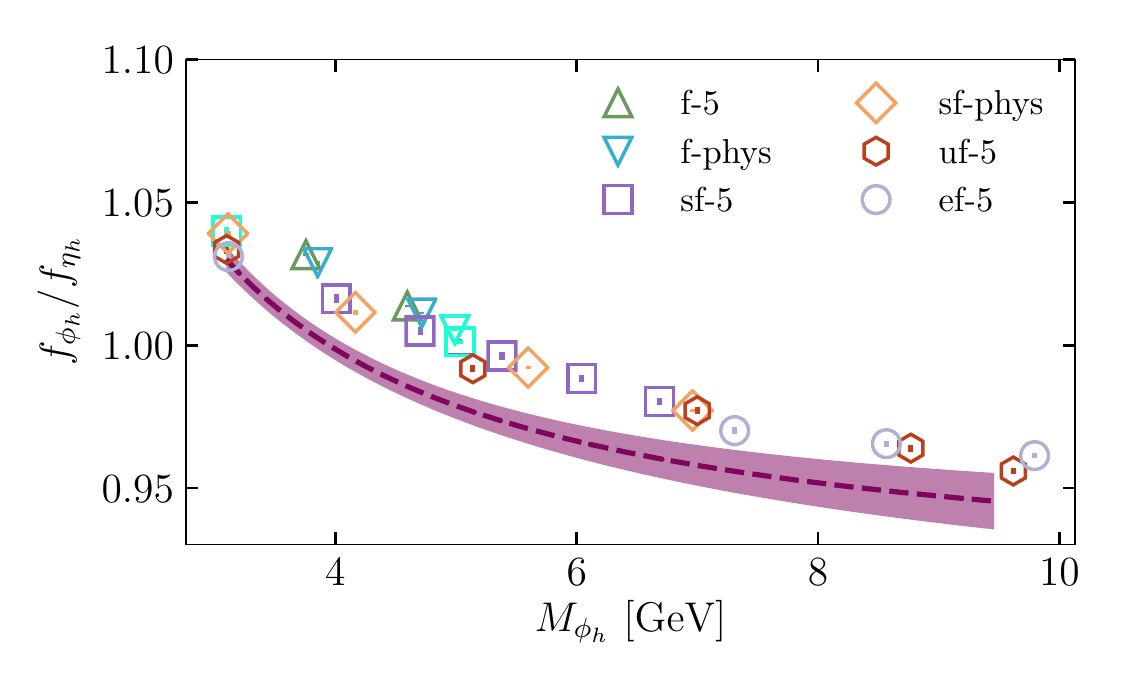}
  \caption{The left-hand plot shows the decay constant of the vector heavyonium meson as a function of its mass from $n_f=2=1+1$ lattice QCD~\cite{Hatton:2021dvg}. The points are the lattice results (with symbols as in Figure~\ref{fig:hyp}) and the purple band shows a model-independent fit using cubic splines. The right-hand plot shows the ratio of vector to pseudoscalar decay constants as a function of the vector heavyonium mass. Some systematic errors, such as discretisation effects, largely cancel in this ratio. This makes it very clear that the ratio changes from being greater than 1 to being less than 1 as the heavy quark mass is increased from $c$ to $b$.}
  \label{fig:fv}
\end{figure}

Figure~\ref{fig:fv} shows the vector meson decay constant and the ratio of vector to pseudoscalar decay constants as a function of the $\phi_h$  mass~\cite{Hatton:2021dvg}. Note that, interestingly, the ratio is larger than 1 at $c$ and falls to less than 1 at $b$. This is unambiguous for our HISQ results because we are able to normalise the currents accurately and fully nonperturbatively. At the $b$ quark mass our results are: $f_{\eta_b}=724(12)$ MeV and 
$f_{\Upsilon} =677.2(9.7)$ MeV.

The result for $f_{\Upsilon}$ can be used to derive a leptonic width for the $\Upsilon$ of $\Gamma(\Upsilon \rightarrow e^+e^-) = 1.292(37)(3) \,\mathrm{keV}$ which can be compared to experiment. Equivalently, Figure~\ref{fig:comp} shows the good agreement between our result for the $\Upsilon$ decay constant and that inferred from the experimental $\Upsilon$ leptonic width.

\section{Results - $m_b$}

\begin{figure}[thb]
  \centering
  \includegraphics[width=0.45\hsize]{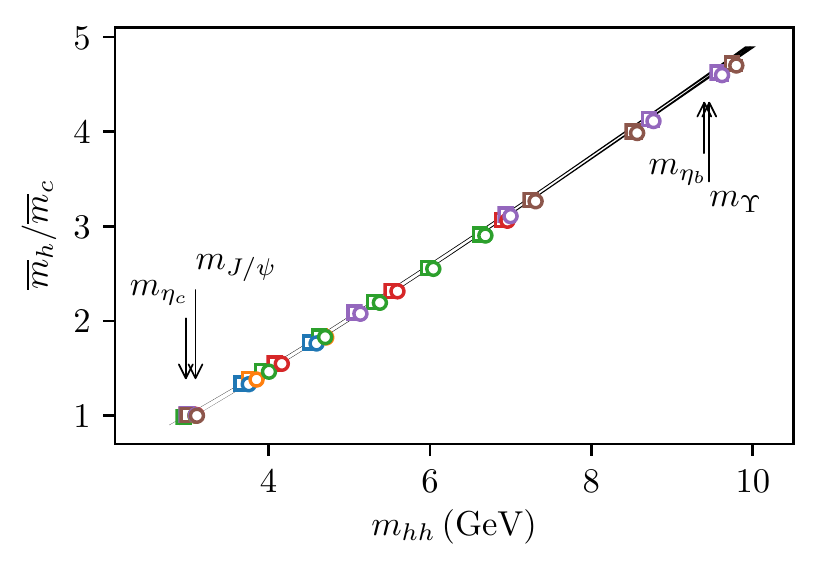}
    \includegraphics[width=0.45\hsize]{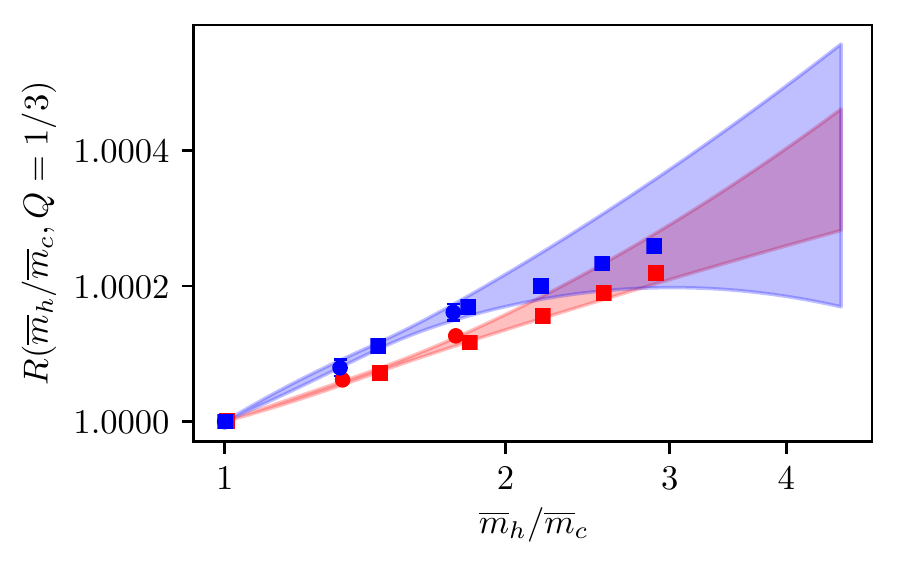}
  \caption{The left-hand plot shows lattice QCD (without QED) results for $m_h/m_c$ plotted agains the heavyonium meson mass. The two lines correspond to using $\eta_h$ and $\phi_h$ and show best fit values for the dependence obtained, using a cubic spline fit~\cite{Hatton:2021syc}. The line thickness shows the $1\sigma$ uncertainty in the fit. The lattice results are shown by squares for the pseudoscalar case (top line) and by circles for the vector case (bottom line). Different colours denote different gluon configurations sets covering the same range as in Figure~\ref{fig:hyp}. The right-hand plot shows the ratio $R$ of $\overline{m}_h/\overline{m}_c$ computed with QED charge $e/3$ to that without QED. The quark masses are tuned to give the same result for heavyonium meson masses, with the red curve using the $\eta_h$ and the blue curve using the $\phi_h$. The blue and red shaded areas show the $\pm 1\sigma$ error bands on the fits to the data~\cite{Hatton:2021syc}. }
  \label{fig:mb}
\end{figure}

\begin{figure}[thb]
  \centering
  \includegraphics[width=0.45\hsize]{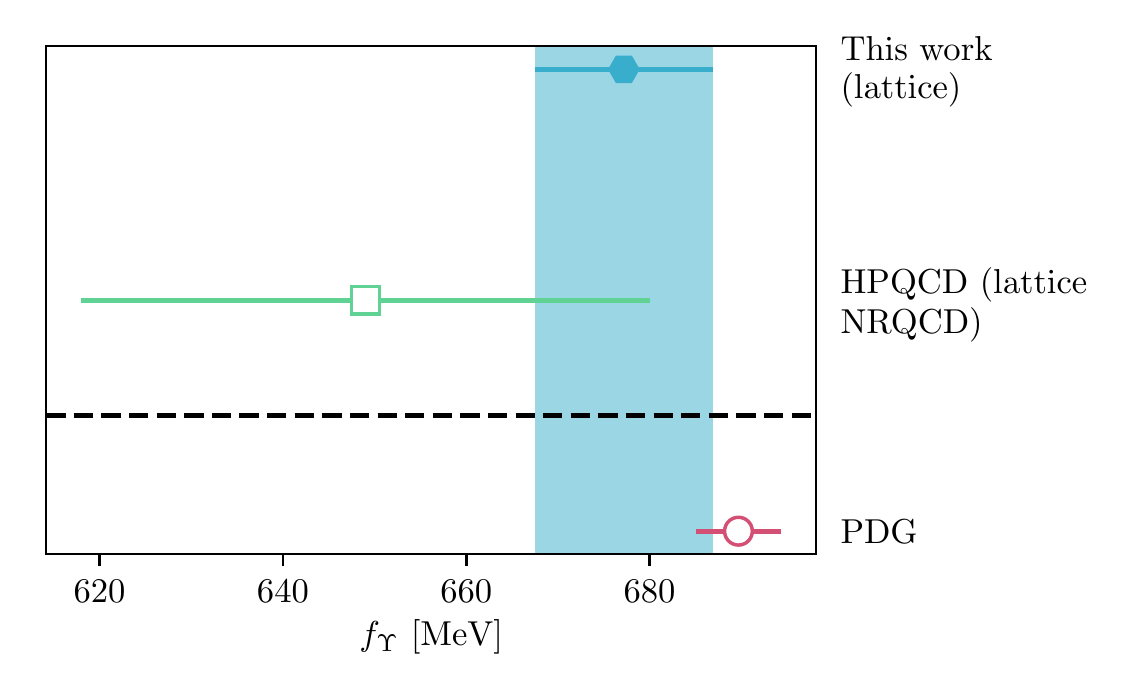}
    \includegraphics[width=0.45\hsize]{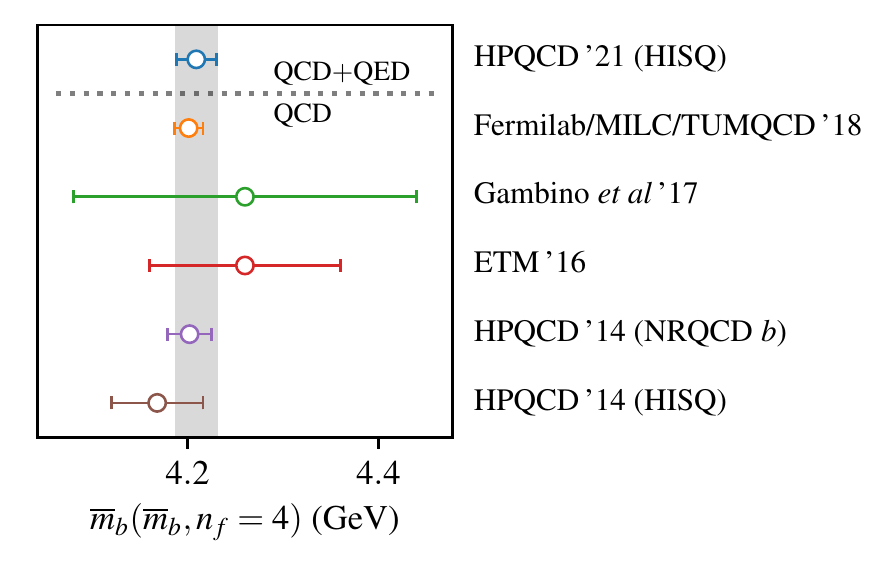}
  \caption{The left-hand plot compares our result (filled blue hexagon) for the $\Upsilon$ decay constant to an earlier HPQCD result using NRQCD $b$ quarks~\cite{Colquhoun:2014ica} (green open square). Our result, carried down as the blue band, is also compared to the experimental average result~\cite{Zyla:2020zbs}, inferred from the $\Upsilon$ leptonic width, given by the pink open circle. We see good agreement with experiment at the level of our 1.4\% uncertainty. The right-hand plot compares our new value for $m_b$ in QCD+QED to earlier results that used lattice QCD calculations, estimating QED effects. These earlier results are from~\cite{FermilabLattice:2018est,Gambino:2017vkx,ETM:2016nbo,Colquhoun:2014ica,Chakraborty:2014aca}.}
  \label{fig:comp}
\end{figure}

Figure~\ref{fig:mb} (left-hand plot) shows the ratio of HISQ masses $m_h/m_c$ as a function of heavyonium meson mass (either $\phi_h$ or $\eta_h$) in pure QCD~\cite{Hatton:2021syc}. Our fit curves (using cubic splines) are also shown. The right-hand plot shows the relative change in this curve if both quarks have electric charge $e/3$. This is a tiny (< 0.04\%) effect. A larger, 0.13\%, QED effect comes from changing the $c$ quark electric charge to $2e/3$. We do this at scale 3 GeV using~\cite{Hatton:2020qhk}. We obtain, for the ratio of $\overline{\text{MS}}$ masses at 3 GeV
\begin{equation}
    \frac{\overline{m}_b(3\,\mathrm{GeV})}{\overline{m}_c(3\,\mathrm{GeV})}\Bigg|_{\substack{\mathrm{QCD} \\ \mathrm{QED}}}
    \!=\,4.586(12).
    \label{eq:mbmcQED}
\end{equation}

We can convert this into a value for $m_b$, using our $m_c$ result in QCD+QED of 0.9841(51) GeV~\cite{Hatton:2020qhk}, giving 
\begin{equation}
    \overline{m}_b(\overline{m}_b)\big|_{\substack{\mathrm{QCD} \\ \mathrm{QED}}} = 
    \begin{cases}
      4.209(21)\,\mathrm{GeV} & n_f=4 \\ 
      4.202(21)\,\mathrm{GeV} & n_f=5, \\ 
    \end{cases}
\end{equation}
from calculations that include QED for the first time~\cite{Hatton:2021syc}. Figure~\ref{fig:comp} compares this to earlier pure QCD results (that estimated QED effects). 

\section{Conclusions}
Following our earlier high precision QCD+QED charmonium calculations using HISQ~\cite{Hatton:2020qhk}, we give here accurate results for $\Upsilon$ and $\eta_b$ properties. Good agreement with experiment is seen, with uncertainties now a few \%~\cite{Hatton:2021dvg}. A new, 0.6\%-accurate, QCD+QED determination of $m_b$ is also given~\cite{Hatton:2021syc}. 

\vspace{2mm}

{\bf Acknowledgements} We are grateful to the MILC collaboration for the use of
their gluon field configurations and for the use of 
MILC's QCD code. 
Computing was done on the Darwin supercomputer at the University of
Cambridge High Performance Computing Service as part of the DiRAC facility,
jointly funded by the Science and Technology Facilities Council,
the Large Facilities Capital Fund of BIS and
the Universities of Cambridge and Glasgow.
We are grateful to the Darwin support staff for assistance.
Funding for this work came from the
Science and Technology Facilities Council
and the National Science Foundation.

%

\bibliography{lat21}

\end{document}